\documentclass{article}

\usepackage{spconf,amsmath,graphicx}
\usepackage{amsmath,graphicx}
\usepackage{amsbsy}
\usepackage{epsfig}
\usepackage{float}
\usepackage{graphicx}
\usepackage{color}
\usepackage{url}
\usepackage{cite}
\usepackage{amssymb}
\usepackage{graphicx}
\usepackage{lipsum}
\usepackage{float}
\usepackage{cuted}
\usepackage{balance}
\usepackage{booktabs}
\usepackage{verbatim}
\usepackage{hyperref}
\usepackage{calc}  
\usepackage{enumitem}  
\usepackage{multirow}
\usepackage{algorithmic}
\usepackage{algorithm2e}
\usepackage{subcaption}
\input{mysymbol.sty}

\ninept
\makeatletter
\newcommand*\dotp{\mathpalette\dotp@{.5}}
\newcommand*\dotp@[2]{\mathbin{\vcenter{\hbox{\scalebox{#2}{$\m@th#1\bullet$}}}}}

\makeatother
\def\x{{\mathbf x}}
\def\f{{\mathbf f}}
\def\g{{\mathbf g}}

\def\F{{\mathbf F}}
\def\U{{\mathbf U}}
\def\G{{\mathbf G}}
\def\D{{\mathbf D}}
\def\S{{\mathbf S}}
\def\h{{\mathbf h}}

\def\tmx{{\mathcal{T}_x\mathcal{M}}}
\def\tmxrp{{\mathcal{T}_x\mathbb{R}^p}}

\def\tm{{\mathcal{T}\mathcal{M}}}
\def\M{{\mathcal{M}}}
\def\i{{\textit{i}}}
\def\rp{{\mathbb{R}^p}}
\def\inc{{\underline{\triangleleft}}}
\def\ltm{{\mathcal{L}^2(\tm)}}
\def\ltmn{{\mathcal{L}^2(\tm_n)}}

\def \Phii{{\boldsymbol{\phi}_i}}
\def \th{{\widetilde{h}}}
\def \bPsi{{\boldsymbol{\Psi}}}
\def\Oi{{\mathbf{O}_i}}
\def\Oj{{\mathbf{O}_j}}
\def\Oij{{\mathbf{O}_{i,j}}}
\def\hd{{\hat{d}}}
\def \samp{\boldsymbol{\Omega}}
\def \tGamma{\widetilde{\Gamma}}
\def \eigGammai{\widetilde{\boldsymbol{\phi}}^n_{i}}
\def \eivGammai{\widetilde{\lambda}^n_{i}}

\title{Tangent Bundle Filters and Neural Networks: \\  from Manifolds to Cellular Sheaves and Back}
\name{C. Battiloro$^{1,2}$, Z. Wang$^1$, H. Riess$^3$, P. Di Lorenzo$^2$, A. Ribeiro$^1$}
\address{
$^1$ ESE Department, University of Pennsylvania, Philadelphia, USA \\
$^2$ DIET Department, Sapienza University of Rome, Rome, Italy \\
$^3$ ECE Department, Duke University, Durham, USA\\
E-mail: claudio.battiloro@uniroma1.it, zhiyangw@seas.upenn.edu
 \vspace{-.25cm}
}
\begin{document}
%
\maketitle
\begin{abstract}
In this work we introduce a convolution operation over the tangent bundle of Riemannian manifolds exploiting the Connection Laplacian operator. We use the convolution to define tangent bundle filters  and tangent bundle neural networks (TNNs), novel continuous architectures operating on tangent bundle signals, i.e. vector fields over manifolds. We  discretize TNNs both in space and time domains, showing that their discrete counterpart is a principled variant of the recently introduced Sheaf Neural Networks. We formally prove that this discrete architecture converges to the underlying continuous TNN. We numerically evaluate the effectiveness of the proposed architecture on a denoising task of a tangent vector field over the unit 2-sphere.
\end{abstract}
\vspace{-.3cm}
\section{Introduction}
\label{sec:intro}
\vspace{-.1cm}
The success of deep learning is mostly the success of Convolutional Neural Networks (CNNs) \cite{lecun1998gradient}. CNNs have achieved impressive performance in a wide range of applications showing good generalization ability. Based on shift operators in the space domain, one (but not the only one) key attribute is that the convolutional filters satisfy the property of shift equivariance. Nowadays, data defined on irregular (non-Euclidean) domains are pervasive, with applications ranging from detection and recommendation in social networks processing\cite{aggarwal2020machine}, to resource allocations over wireless networks \cite{wang2022learning}, or point clouds for shape segmentation \cite{xie2020linking}, just to name a few. For this reason, the notions of shifts in CNNs have been adapted to convolutional architectures on graphs (GNNs) \cite{gama2018convolutional,scarselli2008graph} as well as a plethora of other structures,~e.g.~simplicial complexes \cite{battiloro2022san,bodnar2021weisfeiler,barbarossa2020topological, edgenetsIsufi}, cell complexes \cite{battiloro2022can,bodnarcwnet}, and manifolds \cite{wang2021stability}. In \cite{parada2020algebraic}, a framework for algebraic neural networks has been proposed exploiting commutative algebras. In this work we focus on tangent bundles, a formal tool for describing and processing vector fields on manifolds, which are key elements in tasks such as robot navigation or flocking modeling.

\noindent\textbf{Related Works.}  The renowned manifold assumption states that high dimensional data examples are sampled from a low-dimensional Riemannian manifold. This assumption is the fundamental block of manifold learning, a class of methods for non-linear dimensionality reduction. Some of these methods approximate manifolds with k-NN or geometric graphs via sampling points, i.e., for a fine enough sampling resolution, the graph Laplacian of the approximating graph ``converges'' to the Laplace-Beltrami operator of the manifold \cite{belkin2008towards}. These techniques rely on the eigenvalues and eigenvectors of the graph Laplacian \cite{chung1997spectral}, and they give rise to a novel perspective on manifold learning. In particular, the above approximation leads to  important transferability results of graph neural networks (GNNs) \cite{ruiz2021transferability, levie2019transferability}, as well as to the introduction of Graphon and Manifold Neural Networks, continuous architectures shown to be limit objects of GNNs \cite{wang2022convolution, ruiz2021graphon}. However, most of the previous works focus on scalar signals, e.g. one or more scalar values attached to each node of  graphs or point of  manifolds; recent developments \cite{Sharp2019vhm} show that processing vector data defined on tangent bundles of manifolds or discrete vector bundles \cite{hansen2019sheafsp,bodnar2022sheafdiff} comes with a series of benefits. Moreover, the work in \cite{singer2012vdm} proves that it is possible to approximate both manifolds and their tangent bundles with certain cellular sheaves obtained from a point cloud via k-NN and Local PCA, such that, for a fine enough sampling resolution, the Sheaf Laplacian of the approximating sheaf ``converges'' to the Connection Laplacian operator. Finally, the work in \cite{singer2013spectral} generalizes the result of \cite{singer2012vdm} by proving the spectral convergence of a large class of Laplacian operators via the Principal Bundle set up.

\noindent\textbf{Contributions.} In this work we define a convolution operation over the tangent bundles of Riemannian manifolds with the Connection Laplacian operator. Our definition is consistent, i.e. it reduces to manifold convolution \cite{wang2022convolution} in the one-dimensional bundle case, and to the standard convolution if the manifold is the real line. We introduce tangent bundle convolutional filters to process tangent bundle signals (i.e. vector fields over manifolds), we define a frequency representation for them and, by cascading  layers consisting of tangent bundle filters banks and nonlinearities, we introduce Tangent Bundle Neural Networks (TNNs). We then discretize the TNNs in the space domain by sampling points on the manifold and building a cellular sheaf \cite{Hansen2019towardspecsheaf} representing a legit approximation of both the manifold and its tangent bundle \cite{singer2012vdm}. We formally prove that the discretized architecture over the cellular sheaf converges to the underlying TNN as the number of sampled points increases. Moreover, we further discretize the architecture in the time domain by sampling the filter impulse function in discrete and finite time steps, showing that space-time discretized TNNs are a principled variant of the very recently introduced Sheaf Neural Networks \cite{bodnar2022sheafdiff,hansen2020sheafnn,barbero2022sheafnnconn}, discrete architectures operating on cellular sheaves and generalizing graph neural networks. Finally, we numerically evaluate the performance of TNNs on a denoising task of a tangent vector field of the unit 2-sphere.

\noindent\textbf{Paper Outline.} The paper is organized as follows. We start with some preliminary concepts in Section 2.  We define the tangent bundle convolution and filters in Section 3, and  Tangent Bundle Neural Networks (TNNs) in Section 4. In Section 5, we discretize TNNs in space and time domains, showing that discretized TNNs are Sheaf Neural Networks and proving the convergence result. Numerical results are in Section 6 and conclusions are in Section 7. 
\vspace{-.2cm}
\section{Preliminary Definitions}
\label{subsec:prelim_def}
\vspace{-.1cm}
\noindent\textbf{Manifolds and Tangent Bundles.} We consider a compact and smooth $d-$dimensional manifold $\mathcal{M}$ isometrically embedded in $\mathbb{R}^p$. Each point $x \in \mathcal{M}$ is endowed with a $d-$dimensional tangent (vector) space $\tmx \cong \mathbb{R}^d$, $\mathbf{v} \in \tmx$ is said to be a tangent vector at $x$ and can be seen as the velocity vector of a curve over $\mathcal{M}$ passing through the point $x$ (formal definitions can be found in \cite{lee2006riemannian}). The disjoint union of the tangent spaces is called the tangent bundle $\tm = \bigsqcup_{x \in \M} \tmx$. The embedding induces a Riemann structure on $\mathcal{M}$; in particular, it equips each tangent space $\tmx$ with an inner product, called Riemann metric, given, for each $\mathbf{v}$,$\mathbf{w} \in \tmx$, by
\begin{equation}
\label{riemann_metric}
    \langle\mathbf{v},\mathbf{w}\rangle_{\tmx} = \i\mathbf{v} \dotp \i\mathbf{w},
\end{equation}
where $\i\mathbf{v} \in \tmxrp$ is the embedding of $\mathbf{v} \in \tmx$ in $\tmxrp \subset \rp$ (the d-dimensional subspace of $\rp$ which is the embedding of $\tmx$ in $\rp$), with $\i:\tm\rightarrow\tmxrp$ being an injective linear mapping  referred to as  differential \cite{lee2006riemannian}, and $\dotp$ is the  dot product. The  Riemann metric induces also a probability measure $\mu$ over the manifold.

\noindent \textbf{Tangent Bundle Signals.} A tangent bundle signal is a vector field over the manifold, thus a mapping $\F: \M \rightarrow \tm$ that associates to each point of the manifold a vector in the corresponding tangent space. An inner product for tangent bundle signals $\F$ and $\mathbf{G}$ is
\begin{equation}
\label{inn_prod}
\langle \F, \mathbf{G} \rangle_{\tm} = \int_{\M} \langle \F(x), \mathbf{G}(x) \rangle_{\tmx} \textrm{d}\mu(x),
\end{equation}
and the induced norm is $||\F||^2_{\tm} = \langle \F, \F \rangle_{\tm}$. We denote with $\ltm$ the Hilbert Space of finite energy (w.r.t. $|| \cdot ||_{\tm}$) tangent bundle signals. In the following we denote $\langle \cdot, \cdot \rangle_{\tm}$ with $\langle \cdot, \cdot \rangle$  when there is no risk of confusion.

\noindent \textbf{Connection Laplacian.} The Connection Laplacian is a (second-order) operator $\Delta: \ltm \rightarrow \ltm$, given by the trace of the second covariant derivative defined (for this work) via the Levi-Cita connection \cite{singer2012vdm}. The connection Laplacian $\Delta$ has some desirable properties: it is negative semidefinite, self-adjoint and elliptic. The Connection Laplacian characterizes the heat diffusion equation
\begin{equation}
\label{diff_eq}
\frac{\partial \U(x,t)}{\partial t} - \Delta\U(x,t) = 0,
\end{equation}
where $\U: \M \times \mathbb{R}_0^+ \rightarrow \tm$ and $\U( \cdot, t) \in \ltm \, \forall t \in \mathbb{R}_0^+$ (see \cite{Sharp2019vhm} for a simple interpretation of  \eqref{diff_eq}). With initial condition set as $\U( x, 0) = \F(x)$, the solution of \eqref{diff_eq} is given by
\begin{equation}
\label{exp_sol}
\U( x, t) = e^{t\Delta}\F(x),
\end{equation}
which provides a way to construct tangent bundle convolution, as explained in the following section. The Connection Laplacian $\Delta$ has a negative  spectrum $\{-\lambda_i, \Phii\}_{i=1}^{\infty}$ with eigenvalues $\lambda_i$ and  corresponding eigenvector fields $\Phii$ satisfying
\begin{equation}
\label{eigen}
\Delta\Phii = -\lambda_i\Phii,
\end{equation}
with $0<\lambda_1\leq\lambda_2\leq\dots$. The $\lambda_i$s and the  $\Phii$s can be interpreted as the canonical frequencies and oscillation modes of $\tm$. 
\vspace{-.2cm}
\section{Tangent Bundle Convolutional Fitlers}
\label{sec:filters}
\vspace{-.1cm}
In this section we define the tangent bundle convolution of a filter impulse response $\th$ and a tangent bundle signal $\F$. 

\noindent{\textbf{\textit{Definition 1. (Tangent Bundle Filter)}}} Let $\th:\mathbb{R}^+ \rightarrow \mathbb{R}$ and let $\F \in \ltm$ be a tangent bundle signal. The manifold filter with impulse response $\th$, denoted with $\h$, is given by
\begin{equation}
    \label{convolution}
    \G(x) = \h\F(x) :=\big(\th \star_{\tm} \F\big)= \int_0^{\infty}\th(t)\U(x,t)\textrm{d}t,
\end{equation}
where $\U(x,t)$ is the solution of the heat equation in \eqref{diff_eq} with $\U(x,0) = \F(x)$. Injecting \eqref{exp_sol} in \eqref{convolution}, we obtain
\begin{equation}
    \label{param_conv}
    \G(x) = \h\F(x) = \int_0^{\infty}\th(t)e^{t\Delta}\F(x)\textrm{d}t = \h(\Delta)\F(x).
\end{equation}
The convolution in Definition 1 is consistent, i.e. it generalizes the manifold convolution \cite{wang2022convolution} and the standard convolution in Euclidean domains (see Appendix A.4). The frequency representation $\hat{F}$ of $\F$ can be obtained by projecting $\bbF$ onto the  $\Phii$s basis
\begin{equation}
\label{freq_resp}
    \big[\hat{F}\big]_i = \langle \F, \Phii \rangle = \int_{\M}\langle \F(x), \Phii(x) \rangle_{\tmx} \textrm{d}\mu(x)
\end{equation}

\noindent{\textbf{\textit{Definition 2. (Bandlimited Tangent Bundle Signals)}}} A tangent bundle signal is said to be $\lambda_M$-bandlimited with $\lambda_M>0$ if $\big[\hat{F}\big]_i=0$ for all i such that $\lambda_i > \lambda_M$.

\noindent{\textbf{\textit{Proposition 1.}}} Given a tangent bundle signal $\F$ and a tangent bundle filter $\h(\Delta)$ as in Definition 1, the frequency representation of the filtered signal $\G = \h(\Delta)\F$ is given by
\begin{equation}
    \big[\hat{G}\big]_i = \int_0^{\infty} \th(t)e^{-t\lambda_i}\textrm{d}t\big[\hat{F}\big]_i.
\end{equation}
\noindent{\textbf{\textit{Proof.}}} See Appendix A.1.

\noindent{\textbf{\textit{Definition 3. (Frequency Response)}}}The frequency response $\hat{h}(\lambda)$ of the filter $\h(\Delta)$ is defined as
\begin{equation}
    \label{frequency_filt}
    \hat{h}(\lambda) = \int_0^{\infty} \th(t)e^{-t\lambda}\textrm{d}t.
\end{equation}
This leads to $\big[\hat{G}\big]_i = \hat{h}(\lambda_i)\big[\hat{F}\big]_i$, meaning that the  tangent bundle filter is point-wise in the frequency domain. Therefore, we can write the frequency representation of the tangent bundle filter as
\begin{equation}
\label{filtered_series}
\G = \h(\Delta)\F = \sum_{i=1}^{\infty} \hat{h}(\lambda_i) \langle \F, \Phii \rangle \Phii.
\end{equation}
We note that the frequency response of the tangent bundle filter generalizes the frequency response of a standard time filter as well as a graph filter \cite{shuman2013emerging}. 
\vspace{-.2cm}
\section{Tangent Bundle Neural Networks}\label{sec:tnn}
\vspace{-.1cm}
We define a layer of a Tangent Bundle Neural Network (TNN) as a bank of tangent bundle filters followed by a pointwise non-linearity. In this setting,  pointwise informally means ``pointwise in the ambient space". We introduce the notion of differential-preserving non-linearity to formalize this concept.

\noindent{\textbf{\textit{Definition 4. (Differential-preserving Non-Linearity)}}} Denote with $U_x \subset \tmxrp$ the image of the injective differential $\i$ in $x$. A mapping $\sigma:\ltm \rightarrow \ltm$ is a differential-preserving non-linearity if it can be written as $\sigma(\F(x)) = \i^{-1} \circ \widetilde{\sigma}_x \circ \i\F(x)$, where $\widetilde{\sigma}_x: U _x\rightarrow U_x$ is a point-wise non-linearity in the usual (euclidean) sense. 

Furthermore, we assume that $\widetilde{\sigma}_x = \widetilde{\sigma}$ for all $x \in \M$. Thus, the $l-$th layer of a TNN with $F_l$ input signals $\{\F_l^q\}_{q = 1}^{F_l}$, $F_{l+1}$ output signals $\{\F_{l+1}^u\}_{u = 1}^{F_{l+1}}$,  and point-wise non linearity $\sigma(\cdot)$ is written as
\begin{equation}
    \label{tnn_layer}
    \F_{l+1}^u(x) = \sigma\Bigg(\sum_{q=1}^{F_l}\h(\Delta)_l^{u,q}\F_l^q(x)\Bigg), \; u = 1,...,F_{l+1}.
\end{equation}
A TNN of depth $L$ with input signals $\{\mathbf{F}^q\}_{q=1}^{F_0}$ is built as the stack of $L$ layers defined in \eqref{tnn_layer}, where $\F_0^q = \F^q$. To globally represent the TNN, we collect all the filter impulse responses in a function set $\mathcal{H} = \big\{\hat{h}_l^{u,q}\big\}_{l,u,q}$ and we describe the TNN $u-$th output as a mapping $\F^u_{L}=\bPsi_u\big(\mathcal{H}, \Delta, \{\mathbf{F}^q\}_{q=1}^{F_0}\big)$ to enhance that it is parameterized by filters $\mathcal{H}$ and Connection Laplacian $\Delta$.
\vspace{-.2cm}
\section{Discretization in Space and Time}
\vspace{-.1cm}
Tangent Bundle Filters and Tangent Bundle Neural Networks operate on tangent bundle signals, thus they are continuous architectures that cannot be directly implemented in practice. Here we provide a principled way of discretizing them both in time and space domains.

\noindent\textbf{Discretization in the Space Domain.}
The manifold $\M$, the tangent bundle $\tm$, and the Connection Laplacian $\Delta$ can be approximated starting from a set of sampled points (point-cloud). Knowing the coordinates of the sampled points, it is indeed possible to build a specific (orthogonal) cellular sheaf  over an undirected geometric graph (see Appendix A.3) such that its  Sheaf Laplacian converges to the manifold Connection Laplacian as the number of sampled points (nodes) increases \cite{singer2013spectral}. We assume that a set of $n$  points $\mathcal{X}=\{x_1,\dots,x_n\}\subset \mathbb{R}^p$ are sampled i.i.d. from measure $\mu$ over $\mathcal{M}$. We  build a cellular sheaf $\tm_n$ following the Vector Diffusion Maps procedure whose details are listed in \cite{singer2012vdm}. In particular, we build a geometric graph $\M_n$, with weights  for nodes $i$ and $j$  set as
\begin{equation}
\label{graph_weights} 
    w_{i,j} = \exp\Bigg(\frac{||x_i-x_j||^2}{\sqrt{\epsilon}}\Bigg)\mathbb{I}\Big(0 < ||x_i-x_j||^2\leq \sqrt{\epsilon}\Big),
\end{equation}
where $\epsilon$ controls the chosen Gaussian Kernel. We then assign to each node $i$ an orthogonal transformation $\Oi \in \mathbb{R}^{p\times\hd}$ computed via a local PCA procedure, that is an approximation of a basis of the tangent space $\mathcal{T}_{x_i}\mathcal{M}$, where $\hd$ is an estimate of $d$ obtained from the same procedure. At this point, an approximation of the transport operator \cite{lee2006riemannian} from $\mathcal{T}_{x_i}\mathcal{M}$ to $\mathcal{T}_{x_j}\mathcal{M}$ is also needed. In the discrete domain, this translates in associating a matrix to each edge of the above graph (the restriction maps of the sheaf). For $\epsilon$  small enough, 
$\mathcal{T}_{x_i}\mathcal{M}$ and $\mathcal{T}_{x_j}\mathcal{M}$ are close, meaning that the column spaces of $\Oi$
and $\Oj$ are similiar. If they were coinciding, then
the matrices $\Oi$
and $\Oj$ would have been the same up to an orthogonal transformation $\widetilde{\mathbf{O}}_{i,j}$
satisfying $\widetilde{\mathbf{O}}_{i,j} = \Oi^T\Oj$. However,  the subspaces are not coinciding due to
curvature. For this season, the transport operator approximation $\Oij$ is defined as the closest orthogonal matrix \cite{singer2012vdm} to $\widetilde{\mathbf{O}}_{i,j}$, and it is computed as $\Oij = \mathbf{M}_{i,j}\mathbf{V}^T_{i,j} \in \mathbb{R}^{\hd\times\hd}$, where $\mathbf{M}_{i,j}$ and $\mathbf{V}_{i,j}$ are the SVD of $\widetilde{\mathbf{O}}_{i,j} = \mathbf{M}_{i,j}\boldsymbol{\Sigma}_{i,j}\mathbf{V}^T_{i,j}$. We now build a block matrix $\S \in \mathbb{R}^{n\hd\times n\hd}$ and a diagonal block matrix $\D \in \mathbb{R}^{n\hd\times n\hd}$ with $\hd \times \hd $ blocks defined as
\begin{align}
\label{DS}
&\S_{i,j} = w_{i,j}\widetilde{\mathbf{D}}_i^{-1}\Oij\widetilde{\mathbf{D}}_j^{-1}, \quad\D_{i,i} = \textrm{ndeg(i)}\mathbf{I}_{\hd} ,
\end{align}
where  $\widetilde{\mathbf{D}}_i = \textrm{deg}(i)\mathbf{I}_{\hd}$, $\textrm{deg}(i) = \sum_{j}w_{i,j}$ is the degree of node $i$, and $\textrm{ndeg}(i) = \sum_{j}w_{i,j}/(\textrm{deg}(i)\textrm{deg}(j))$. Finally, we define the (normalized) Sheaf Laplacian as the following matrix
\begin{equation}
\label{sheaf_laplacian}
    \Delta_n = \epsilon^{-1}\big(\D^{-1}\S - \mathbf{I}\big) \in \mathbb{R}^{n\hd\times n\hd},
\end{equation}
which is the approximated Connection Laplacian of the discretized manifold. A sheaf $\tm_n$ with this (orthogonal) structure is also said to be a discrete $\mathcal{O}\big(\hd\big)-$bundle and represents a discretized version of  $\tm$. We introduce a linear sampling operator $\samp_n^{\mathcal{X}}:\ltm \rightarrow \ltmn$ to discretize a  tangent bundle signal $\F$ as a sheaf signal $\f_n \in \mathbb{R}^{n\hd}$ (a 0-cochain of the sheaf)  such that
\begin{align}
\label{sampler}
    &\f_n = \samp_n^{\mathcal{X}}\F, \\
    &\f_{n}(x_i):=[\f_n]_{((i-1)\hd+1):(i+1)\hd} = \Oi^T\i\F(x_i).
\end{align}
We are now in the condition of plugging the discretized operator and signal in the definition of tangent bundle filter in \eqref{param_conv}, obtaining
\begin{equation}
    \label{discr_param_conv}
    \g_n = \int_0^{\infty}\th(t)e^{\Delta_n}\f_n\textrm{d}t = \h(\Delta_n)\f_n \in \mathbb{R}^{n\hd}.
\end{equation}
Following the same considerations of Section \ref{sec:tnn}, we can define a  discretized space tangent bundle neural network (D-TNN) as the stack of $L$ layers of the form
\begin{equation}
    \label{dt_tnn_layer}
    \x_{n,l+1}^u = \sigma\Bigg(\sum_{q=1}^{F_l}\h(\Delta_n)_l^{u,q}\x_{n,l}^q\Bigg), \; u = 1,...,F_{l+1},
\end{equation}
where (with a slight abuse of notation) $\sigma$ has the same point-wise law of $\widetilde{\sigma}$ in Definition 4.  As in the continuous case, we describe the $u-th$ output of a D-TNN as a mapping $\bPsi_u\big(\mathcal{H}, \Delta_n, \{\x_n^q\}_{q=1}^{F_0}\big)$ to enhance that it is parameterized by filters $\mathcal{H}$ and the Sheaf Laplacian $\Delta_n$. As the number of sampling points goes to infinity, the Sheaf Laplacian $\Delta_n$ converges to the Connection Laplacian $\Delta$ and the  sheaf signal $\x_n$ converges to the tangent bundle signal $\F$. Combining these results, we prove in the next proposition that the output of a D-TNN converges to the output of the corresponding TNN as the sample size increases.

\noindent\textbf{\textit{Theorem 1.}} Let $\mathcal{X}=\{x_1,\dots,x_n\}\subset \mathbb{R}^p$ be a set of $n$ i.i.d. sampled points from measure $\mu$ over $\M \subset \mathbb{R}^p$ and $\F$ a bandlimited tangent bundle signal. Let $\tm_n$ be a cellular sheaf built from $\mathcal{X}$ as explained above, with $\epsilon = n^{-2/(\hd+4)}$. Let  $\bPsi_u\big(\mathcal{H}, \cdot, \cdot \big)$ be the $u-th$ output of a neural network with $L$ layers parameterized
by the operator $\Delta$ of  $\tm$  or by the discrete operator $\Delta_n$ of  $\tm_n$. If:
\begin{itemize}
    \item the frequency response of filters in $\mathcal{H}$ are non-amplifying Lipschitz continuous;
    \item the non-linearities are differential-preserving;
    \item $\widetilde{\sigma}$ from Definition 4 is point-wise normalized Lipschitz continuous,
    \item $\samp_n^\mathcal{X} \F$ is a bandlimited sheaf signal
\end{itemize}
then it holds for each $u = 1, 2, \dots, F_L$ that:
\begin{equation}
\label{convergence}
\lim_{n \rightarrow \infty} ||\bPsi_u\big(\mathcal{H}, \Delta_n, \samp_n^{\mathcal{X}}\F\big) - \samp_n^{\mathcal{X}}\bPsi_u\big(\mathcal{H}, \Delta,\F\big)||_{\tm_n} = 0,
\end{equation}
with the limit taken in probability.

\noindent\textbf{\textit{Proof.}} See Appendix A.2.
\begin{table*}[]
\centering\begin{tabular}{@{}lcccc@{}}
\cmidrule(l){3-5}
                                               &                             & $ \sigma = 10^{-2}$                                                    & $\sigma = 5 \cdot 10^{-2}$                                            & $\sigma = 1\cdot 10^{-1}$                                               \\ \midrule
\multicolumn{1}{|l|}{\multirow{2}{*}{$n=200$}} & \multicolumn{1}{c|}{DD-TNN} & \multicolumn{1}{c|}{$\mathbf{2\cdot 10^{-4}} \pm 1.6\cdot 10^{-5}$}   & \multicolumn{1}{c|}{$\mathbf{4.9\cdot 10^{-3}} \pm 2.4 \cdot 10^{-4}$}  & \multicolumn{1}{c|}{$\mathbf{1.9\cdot 10^{-2}} \pm 1.3 \cdot 10^{-3}$}    \\
\multicolumn{1}{|l|}{}                         & \multicolumn{1}{c|}{MNN}    & \multicolumn{1}{c|}{$2.9\cdot 10^{-4} \pm 1.5 \cdot 10^{-5}$}         & \multicolumn{1}{c|}{$7\cdot 10^{-3} \pm 2.8 \cdot 10^{-4}$}         & \multicolumn{1}{c|}{$2.9\cdot 10^{-2} \pm 1.5 \cdot 10^{-3}$}           \\ \midrule
\multicolumn{1}{|l|}{\multirow{2}{*}{$n=800$}} & \multicolumn{1}{c|}{DD-TNN} & \multicolumn{1}{c|}{$\mathbf{2\cdot 10^{-4}} \pm 5.7\cdot 10^{-6}$}   & \multicolumn{1}{c|}{$\mathbf{5\cdot 10^{-3}} \pm 1.2  \cdot 10^{-4}$} & \multicolumn{1}{c|}{$\mathbf{1.9 \cdot 10^{-2}} \pm 4.6 \cdot 10^{-4}$} \\
\multicolumn{1}{|l|}{}                         & \multicolumn{1}{c|}{MNN}    & \multicolumn{1}{c|}{$2.8\cdot 10^{-4} \pm 8.7 \cdot 10^{-6}$} & \multicolumn{1}{c|}{$7.3\cdot 10^{-3} \pm 1.7 \cdot 10^{-4}$}         & \multicolumn{1}{c|}{$2.9\cdot 10^{-2} \pm 6.9 \cdot 10^{-4}$}           \\ \bottomrule
\end{tabular}
\caption{MSE on the denoising task}
     \label{table:results}
\end{table*}

\noindent\textbf{Discretization in the Time Domain.} The discretization in space introduced in the previous section is still not enough for implementing TNNs in practice. Indeed, from Definition 1, we should learn the continuous time function $\tilde{h}(t)$, and this  is generally infeasible. To make TNNs and their training implementable, we discretize the function $\tilde{h}(t)$ in the continuous time domain with a fixed sampling interval $T_s$. We replace the filter response function with a series of coefficients $h_k = \tilde{h}(k T_s)$, $k =0 ,1, 2\dots$. With $T_s=1$ and fixing $K$ samples over the time horizon, the discrete-time version of the convolution in \eqref{convolution} can be thus written as
\begin{equation}
\label{eqn:manifold_convolution_discrete}
    \bbh(\Delta_n) \F(x)= \sum_{k=0}^{\infty} h_k e^{k\Delta}\F(x), 
\end{equation}
which corresponds to the form of a finite impulse response (FIR) filter with shift operator $e^{\Delta}$. We can now
 inject the space discretization in the finite-time architecture in \eqref{eqn:manifold_convolution_discrete}, obtaining an implementable manifold filter on the discretized manifold (cellular sheaf) $\tm_n$ as
\begin{equation}
\label{eqn:discrete_manifold_convolution_discrete}
  \mathbf{g}_n=  \bbh(\Delta_n) \f_n = \sum_{k=0}^{K-1} h_k e^{k\Delta_n}\f_n.
\end{equation}
The discretized manifold filter of order $K$ can be seen as a generalization of graph convolution \cite{gama2018convolutional} to the (orthogonal) cellular sheaf domain, thus we refer $e^{\Delta_n}$ as a sheaf shift operator. At this point, by replacing the filter $\bbh_l^{pq}(\Delta_n)$ in \eqref{dt_tnn_layer} with \eqref{eqn:discrete_manifold_convolution_discrete}, we obtain the following architecture:
\begin{equation}
    \label{sheaf_nn_layer}
    \x_{n,l+1}^u = \sigma\Bigg(\sum_{q=1}^{F_l}\sum_{k=1}^{K}h_{k,l}^{u,q}\big(e^{\Delta_n}\big)^k\x_{n,l}^q\Bigg), \; u = 1,...,F_{l+1},
\end{equation}
that we refer to as  discretized space-time tangent bundle neural network (DD-TNN), which can be seen as a principled variant of the recently proposed Sheaf Neural Networks \cite{hansen2020sheafnn,bodnar2022sheafdiff,barbero2022sheafnnconn}, with $e^{\Delta_n}$  as (sheaf) shift operator with order $K$ diffusion. The layer in \eqref{sheaf_nn_layer} can be rewritten in  matrix form by introducing the matrices $\mathbf{X}_{n,l}=\{\x_{n,l}^u\}_{u=1}^{F_{l}}\in \mathbb{R}^{n\hd \times F_{l}}$, and $\mathbf{H}_{l,k} = \{h_{k,l}^{u,q}\}_{q=1,u=1}^{F_l,F_{l+1}}\in \mathbb{R}^{F_l \times F_{l+1}}$ as
\begin{equation}
    \mathbf{X}_{n,l+1} = \sigma\Bigg(\sum_{k=1}^{K}\big(e^{\Delta_n}\big)^k\mathbf{X}_{n,l}\mathbf{H}_{l,k}\Bigg) \; \in \mathbb{R}^{n\hd \times F_{l+1}},
\end{equation}
where the filter weights $\{\mathbf{H}_{l,k}\}_{l,k}$ are learnable parameters. We have completed the process of building TNNs from cellular sheaves and back. Manifolds and their Tangent Bundles can be seen as the limits of graphs and  cellular sheaves on them, making TNNs also a tool for analyzing large graphs with vector data.
\vspace{-.2cm}
\section{Numerical Results}
\vspace{-.1cm}
We assess the consistency of the proposed framework by designing a denoising task\footnote[1]{\scriptsize  \url{https://github.com/clabat9/Tangent-Bundle-Neural-Networks}}. We work on the unit 2-sphere ($\ccalM=\mathcal{S}_2$) and its tangent bundle. In particular, we uniformly sample the sphere on $n$ points $\mathcal{X}=\{\x_1,\dots,\x_n\}$, and we compute the corresponding cellular sheaf $\tm_n$, Sheaf Laplacian $\Delta_n$ and signal sampler $\samp_n^{\mathcal{X}}$ as explained in Section 5 (also obtaining $\hd=2$). We consider the tangent vector field over the sphere given by
\begin{equation}
\i\F(x,y,z)=(-y,x,0) \in 
\mathbb{R}^3,
\end{equation} 
depicted in Fig. 1 for a realization of $\mathcal{X}$ with $n=200$. At this point, we add AWGN with variance $\tau^2$ to $\i\F$ obtaining a noisy field $\widetilde{\i\F}$, then we use $\samp_n^{\mathcal{X}}$ to sample it, obtaining $\widetilde{\mathbf{f}}_n \in \mathbb{R}^{2n}$. We test the perfomance of the TNN architecture (implemented with a DD-TNN as in \eqref{sheaf_nn_layer}) by evaluating its ability of denoising $\widetilde{\mathbf{f}}_n$. We exploit a one layer architecture with  $1$ output feature (the denoised signal), and 5 filter taps. We train the architecture to minimize the MSE $\frac{1}{n}\|\widetilde{\mathbf{f}}_n - \mathbf{f}_{n,1}\|^2$ between the noisy signal $\widetilde{\mathbf{f}}_n$ and the output of the network $\mathbf{f}_{n,1}$ via the ADAM optimizer \cite{kingma2014adam}, with hyperparameters set to obtain the best results. We compare our architecture with a 1-layer Manifold Neural Network (MNN) architecture (implemented via a GNN as explained in \cite{wang2022convolution}); to make the comparison fair, $\widetilde{\i\F}$ evaluated on $\mathcal{X}$ is given as input to the MNN, organizing it in a  matrix $\widetilde{\F}_n \in \mathbb{R}^{n \times 3}$. We train the MNN to minimize the MSE $\frac{1}{n}\|\widetilde{\F}_n - \mathbf{F}_{n,1}\|_F^2$, where $\|\|_F$ is the Frobenius Norm and $\mathbf{F}_{n,1}$ is the network output. It is easy to see that the ``two" MSEs used for TNN and MNN are completely equivalent due to the orthogonality of the projection matrices $\Oi$. In Table 1 we evaluate TNNs and MNNs for two different sample sizes ($n = 200$ and $n=800$), for three different noise standard deviation ($\tau = 10^{-2}$,$\tau = 5\cdot10^{-2}$ and $\tau = 10^{-1}$), showing the (again equivalent) MSEs $\frac{1}{n}\|\mathbf{f}_n - \mathbf{f}_{n,1}\|^2$ and $\frac{1}{n}\|\F_n - \mathbf{F}_{n,1}\|_F^2$, where $\mathbf{f}_n$ is the sampling via $\samp_n^{\mathcal{X}}$ of the clean field and $\mathbf{F}_n$ is the matrix collecting the clean field evaluated on $\mathcal{X}$. The results are averaged over 5 sampling realizations and 5 noise realizations per each of them. As the reader can notice from Table 1, TNNs always perform better than MNNs, due to their ``bundle-awareness". Moreover, the mean performance remains stable as the number of points decreases, but the variances increase, meaning that having more sampling points(thus a better estimation of the Connection Laplacian) results in a more stable decision of the network. 
\begin{figure}[t!]
\label{sphere_vf}
\centering
\includegraphics[width=0.42\textwidth]{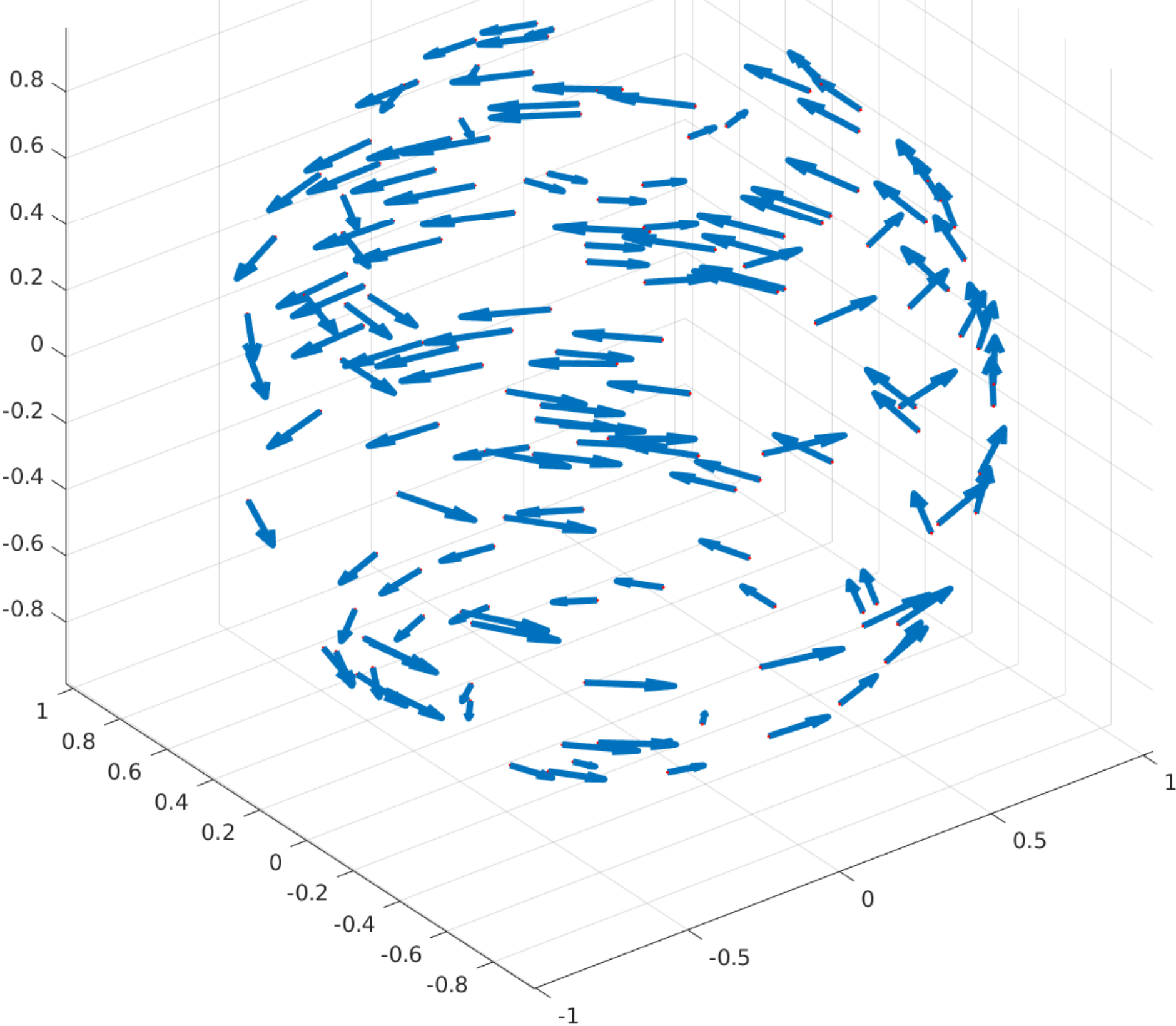}
\caption{Visualization of the embedded tangent vector field $\i\F$} 
\end{figure} 

\vspace{-0.7cm}
\section{Conclusions}
\vspace{-.2cm}
In this work we introduced Tangent Bundle Filters and Tangent Bundle Neural Networks (TNNs), novel continuous architectures operating on tangent bundle signals, i.e. manifold vector fields. We made TNNs implementable by discretization in space and time domains, showing that their discrete counterpart is a principled variant of Sheaf Neural Networks. The results of this preliminary work, in addition to the introduction of a novel tool for processing manifold vector fields, could lead to a deeper  understanding of topological neural architectures in terms of transferability and stability, with the opportunity of designing proper signal processing frameworks on tangent bundles and cellular sheaves. We plan to investigate these problems as well as applying TNNs to real-world complex tasks.

\bibliographystyle{IEEEtran}
\bibliography{refs}

\newpage
\appendix
\section{Appendix}
\subsection{Proof of Proposition 1}
\noindent{\textbf{\textit{Proposition 1.}}} Given a tangent bundle signal $\F$ and a tangent bundle filter $\h(\Delta)$ as in Definition 1, the frequency representation of the filtered signal $\G = \h(\Delta)\F$ is given by:
\begin{equation}
    \big[\hat{G}\big]_i = \int_0^{\infty} \th(t)e^{-t\lambda_i}\textrm{d}t\big[\hat{F}\big]_i.
\end{equation}
\noindent{\textbf{\textit{Proof.}}} By definition of frequency representation in \eqref{freq_resp} we have:
\begin{align}
\label{g_freq}
    &\big[\hat{G}\big]_i = \langle \G, \Phii \rangle = \int_{\M}\langle \G(x), \Phii(x) \rangle_{\tmx} \textrm{d}\mu(x)
\end{align}
Injecting \eqref{param_conv} in \eqref{g_freq}, we get:
\begin{align}
\label{g_freq_f}
    &\big[\hat{G}\big]_i = \langle \int_0^{\infty}\th(t)e^{t\Delta}\F(x)\textrm{d}t, \Phii \rangle 
\end{align}
For the linearity of integrals and inner products, we can write:
\begin{align}
\label{g_freq_int_out}
    &\big[\hat{G}\big]_i = \int_0^{\infty}\th(t)\langle e^{t\Delta}\F(x), \Phii \rangle \textrm{d}t
\end{align}
Finally, exploiting first the self-adjointness of $\Delta$ and then the eigenvector fields definition in \eqref{eigen}, we can write:
\begin{align}
\label{g_freq_final}
    \big[\hat{G}\big]_i &= \int_0^{\infty}\th(t)\langle e^{t\Delta}\F(x), \Phii \rangle \textrm{d}t \nonumber\\
    &=\int_0^{\infty}\th(t)\langle \F(x), e^{t\Delta}\Phii \rangle \textrm{d}t \nonumber \\  
    &=\int_0^{\infty}\th(t)\langle \F(x), e^{-t\lambda_i}\Phii \rangle \textrm{d}t \nonumber \\
    &=\int_0^{\infty}\th(t)e^{-t\lambda_i}\langle \F(x), \Phii \rangle \textrm{d}t,
\end{align}
which concludes the proof.
\subsection{Proof of Theorem 1}
\noindent\textbf{\textit{Theorem 1.}} Let $\mathcal{X}=\{x_1,\dots,x_n\}\subset \mathbb{R}^p$ be a set of $n$ i.i.d. sampled points from measure $\mu$ over $\M \subset \mathbb{R}^p$ and $\F$ a bandlimited tangent bundle signal. Let $\tm_n$ be a cellular sheaf built from $\mathcal{X}$ as explained above, with $\epsilon = n^{-2/(\hd+4)}$. Let  $\bPsi_u\big(\mathcal{H}, \cdot, \cdot \big)$ be the $u-th$ output of a neural network with $L$ layers parameterized
by the operator $\Delta$ of  $\tm$  or by the discrete operator $\Delta_n$ of  $\tm_n$. If:
\begin{itemize}
    \item the frequency response of filters in $\mathcal{H}$ are non-amplifying Lipschitz continuous;
    \item the non-linearities are differential-preserving;
    \item $\widetilde{\sigma}$ from Definition 4 is point-wise normalized Lipschitz continuous,
    \item $\samp_n^\mathcal{X} \F$ is a bandlimited sheaf signal
\end{itemize}
then it holds for each $u = 1, 2, \dots, F_L$ that:
\begin{equation}
\label{convergence}
\lim_{n \rightarrow \infty} ||\bPsi_u\big(\mathcal{H}, \Delta_n, \samp_n^{\mathcal{X}}\F\big) - \samp_n^{\mathcal{X}}\bPsi_u\big(\mathcal{H}, \Delta,\F\big)||_{\tm_n} = 0,
\end{equation}
with the limit taken in probability.

\noindent\textbf{\textit{Proof.}} We define an inner product for sheaf signals $\mathbf{f}$ and $\mathbf{u}$ on a general cellular sheaf $\tm_n$ as:
\begin{align}
\label{emp_metr_sheaf}
\langle \mathbf{f}, \mathbf{u} \rangle_{\tm_n}   & = \frac{1}{n}\sum_{i = 1}^n  \f_{n}(x_i) \dotp \mathbf{u}_{n}(x_i) ,
\end{align}
and the induced norm $||\mathbf{f}||^2_{\tm_n} = \langle \mathbf{f}, \mathbf{f} \rangle_{\tm_n}$.
Under the assumption that the points in $\mathcal{X}$ are sampled i.i.d. from the uniform probability measure $\mu$ given by the induced metric on $\M$ and that $\tm_n$ is built as in Section 5, the inner product in \eqref{emp_metr_sheaf} is equivalent to the following inner product for tangent bundle signals $\F$ and $\U$:
\begin{align}
\label{emp_metr}
\langle \F, \U \rangle_{\tm_n}   &= \int_{\M} \langle \F(x), \U(x) \rangle_{\tmx} \textrm{d}\mu_n(x) \nonumber \\
&= \frac{1}{n}\sum_{i = 1}^n \langle \F(x_i), \mathbf{U}(x_i) \rangle_{\mathcal{T}_{x_i}\M},
\end{align}
and the induced norm $||\F||^2_{\tm_n} = \langle \F, \F \rangle_{\tm_n}$, where $\mu_n = \frac{1}{n}\sum_{i=1}^n\delta_{x_i}$ is the empirical measure corresponding to $\mu$. Indeed, from \eqref{riemann_metric} and due to the orthogonality of the transformations $\Oi$ in Section 5, \eqref{emp_metr} can be  rewritten as:
\begin{align}
\label{emp_metr_versions}
\langle \F, \U \rangle_{\tm_n}&=\frac{1}{n}\sum_{i = 1}^n \langle \F(x_i), \mathbf{U}(x_i) \rangle_{\mathcal{T}_{x_i}\M} \nonumber \\
&= \frac{1}{n}\sum_{i = 1}^n \i\F(x_i) \dotp  \i\mathbf{U}(x_i) \nonumber \\
&= \frac{1}{n}\sum_{i = 1}^n \Oi^T\i\F(x_i) \dotp  \Oi^T\i\mathbf{U}(x_i) \nonumber \\
&= \frac{1}{n}\sum_{i = 1}^n \f_{n}(x_i) \dotp \mathbf{u}_{n}(x_i)  = \langle \mathbf{f}_n, \mathbf{u}_n \rangle_{\tm_n}
\end{align}
where $\f_n = \samp_n^{\mathcal{X}}\F$ and $\mathbf{u}_n = \samp_n^{\mathcal{X}}\U$, respectively. We denote with $\ltmn$ the Hilbert Space of finite energy tangent bundle signals w.r.t. the empirical measure $\mu_n$ (or, equivalently, the Hilbert Space of finite energy sheaf signals w.r.t the norm induced by \eqref{emp_metr_sheaf}). In the following, we will denote the norm $||\cdot||_{\tm_n}$ with $||\cdot||$  when there is no risk of confusion.
We now define bandlimited sheaf signals, Lipshitz continous tangent bundle filters and non-amplifying tangent bundle filters.

\noindent{\textbf{\textit{Definition 5. (Bandlimited Cellular Sheaf Signals)}}} A sheaf signal $\mathbf{f}$ is said to be $M$-bandlimited with $M>0$ if $\big[\hat{f}\big]_i= \langle \mathbf{f}, \Phii^n \rangle_{\tm_n}\neq 0$ only for $i \in \mathcal{F}\subset \{1,\dots,n\}$, with $|\mathcal{F}|=M$ and $\Phii^n$ being the $i$-th eigenvector of the Sheaf Laplacian $\Delta_n$.

\noindent\textbf{\textit{Definition 6. (Tangent Bundle Filters with Lipschitz Continuity)}}
A tangent bundle filter is $C$-Lispchitz if its frequency response is Lipschitz continuous with constant $C$, i.e,
\begin{equation}
    |\hat{h}(a)-\hat{h}(b)| \leq C |a-b|\text{ for all } a,b\in (0,\infty)\text{.}
\end{equation}

\noindent\textbf{\textit{Definition 7. (Non-Amplifying Tangent Bundle Filters)}} A tangent bundle filter is non-amplifying if for all $\lambda\in(0,\infty)$, its frequency response $\hat{h}$ satisfies $|\hat{h}(\lambda)|\leq 1$.

The non-amplifying assumption is reasonable, because the filter function $\hhath(\lambda)$ can always be normalized. In \cite{singer2013spectral}, the spectral convergence of the constructed Sheaf Laplacian in \eqref{sheaf_laplacian} based on the discretized manifold to the Connection Laplacian of the underlying manifold has been proved, and we will exploit that result for proving the following proposition.

\noindent\textbf{\textit{Proposition 3. (Consequence of Theorem 6.3 \cite{singer2013spectral})}}
 Let $\mathcal{X}=\{x_1,\dots,x_n\}\subset \mathbb{R}^p$ be a set of $n$ i.i.d. sampled points from measure $\mu$ over $\M \subset \mathbb{R}^p$. Let $\tm_n$ be a cellular sheaf built from $\mathcal{X}$ as explained in Section 5, with $\epsilon = n^{-2/(\hd+4)}$. Let $\Delta_n$ be the Sheaf Laplacian of $\tm_n$ and $\Delta$ be the Connection Laplacian operator of $\M$. Let $\lambda_{i}^n$ be the $i$-th eigenvalue of $\Delta_n$ and $\Phii^n$ the corresponding eigenvector. Let $\lambda_i$ be the $i$-th eigenvalue of $\Delta$ and $\Phii$  the corresponding eigenvector field of $\Delta$, respectively. Then it holds:
\begin{equation}
\label{eqn:convergence_spectrum}
    \lim_{n\rightarrow \infty } \lambda_i^n = \lambda_i, \quad\lim_{n\rightarrow \infty} \|\Phii^{n} -  \samp_n^{\mathcal{X}}\Phii\|_{\tm_n}=0,
\end{equation}
where the limits are taken in probability.

\noindent\textbf{\textit{Proof.}} These proposition is a consequence of Theorem 6.3 in \cite{chung1997spectral}. Indeed, we  rely on the operator introduced in Definition 6.1 in \cite{singer2013spectral} with $\alpha=1$ and $h_n = n^{-2/(\hd+4)}$ (our $\epsilon$), here denoted as $\Gamma:\ltm \rightarrow \ltm$, and on the operator $\widetilde{\Gamma} = \epsilon^{-1}\big(\Gamma - \textrm{id}\big)$, where $\textrm{id}$ is the identity mapping. It is straightforward to check that:
\begin{equation}
    \label{omega_delta_equiv}
    \widetilde{\Gamma}\F(x_j) = \i^{-1}\Oj\big(\Delta_n\samp_n^{\mathcal{X}}\F\big)(x_j),
\end{equation}
for $j = 1,\dots,n$. We now show that the eigenvectors sampled on $\mathcal{X}$ and eigenvalues of $\widetilde{\Gamma}$  correspond to the eigenvectors and eigenvalues of $\Delta_n$. Let us denote the the $i-th$ eigenvector and eigenvalue of $\widetilde{\Gamma}$ with $\widetilde{\boldsymbol{\phi}}^n_{i}$ and $-\widetilde{\lambda}^n_{i}$, respectively. We have:
\begin{align}
    \label{eigen_conv}
    \tGamma\eigGammai(x_j) = -\eivGammai\eigGammai(x_j)=i^{-1}\Oj\big(\Delta_n\samp_n^{\mathcal{X}}\eigGammai\big)(x_j)
\end{align}
If we apply the mapping $i$ to the last two equalities of \eqref{eigen_conv} and we exploit the orthoghonality of $\Oj$, we obtain:
\begin{align}
\label{eig_gamma_delt}
    \big(\Delta_n\samp_n^{\mathcal{X}}\eigGammai\big)(x_j) = -\eivGammai\Oj^T\i\eigGammai &= -\eivGammai\samp_n^{\mathcal{X}}\eigGammai(x_j)
\end{align}
where the second equality applies the definition of $\samp_n^{\mathcal{X}}$ in \eqref{sampler}. Therefore, we have:
\begin{align}
\label{eig_gamma_delta}
    \lambda_i^n = \eivGammai, \quad \Phii^n(x_j) = \samp_n^{\mathcal{X}}\eigGammai(x_j),
\end{align}
$j= 1,\dots,n$. At this point, we can recall Theorem 6.3 in \cite{singer2013spectral}, that, in the setting of our Theorem 1, states:
\begin{equation}
    \label{spect_VDM}
    \lim_{n\rightarrow \infty } \widetilde{\lambda}_i^n = \lambda_i, \quad\lim_{n\rightarrow \infty} \|\eigGammai -  \Phii\|_{\tm}=0,
\end{equation}
with the limit taken in probability, $i= 1,\dots,n$. 
Injecting the empirical measure in \eqref{spect_VDM} and exploiting the results in \eqref{emp_metr_versions} and \eqref{eig_gamma_delta}, we obtain:
\begin{align}
\label{norm_tm}
    &\|\eigGammai-  \Phii\|_{\tm_n} = \|\Phii^n- \samp_n^{\mathcal{X}}\Phii\|_{\tm_n}
\end{align}
The results in \eqref{spect_VDM} and \eqref{norm_tm}
combined with the a.s. convergence of the empirical measure $\mu_n$ to the measure $\mu$ conclude the proof.

For the sake of clarity, in the following we will drop the dependence on the NNs output index $u$; from the definitions of TNNs in \eqref{tnn_layer} and D-TNNS in \eqref{dt_tnn_layer}, we can thus write:
 \begin{align}
    \nonumber  \|\bPsi\big(\mathcal{H}, \Delta_n, \samp_n^{\mathcal{X}}\F\big) - \samp_n^{\mathcal{X}}\bPsi\big(\mathcal{H}, \Delta,\F\big)\|&= \left\| \bbx_{n,L}-\samp_n^{\mathcal{X}}\F_L \right\|.
 \end{align}
 Further explicating the layers definitions, at layer $l$ we have: 
 \begin{align}
   \nonumber  &\left\| \bbx_{n,l}- \samp_n^{\mathcal{X}} \F_l \right\|\\
     &=\left\| \sigma\left(\sum_{q=1}^{F_{l-1}} \bbh_l^{q}(\Delta_n) \bbx_{n,l-1}^q \right) -\samp_n^{\mathcal{X}} \sigma\left(\sum_{q=1}^{F_{l-1}} \bbh_l^{q}(\Delta) \F_{l-1}^q\right) \right\|
 \end{align}
 with $\bbx_{n,0}^q=\samp_n^{\mathcal{X}} \F^q$ for $q=1,\dots,F_0$. Exploiting the normalized point-wise Lipschitz continuity of the non-linearities and the linearity of the sampling operator $\samp_n^{\mathcal{X}}$, we have:
  \begin{align}
  \label{proof_1}
    \| \bbx_{n,l} - \samp_n^{\mathcal{X}} \F_l  \| &\leq \left\|  \sum_{q=1}^{F_{l-1}} \bbh_l^{q}(\Delta_n) \bbx_{n,l-1}^q    - \samp_n^{\mathcal{X}} \sum_{q=1}^{F_{l-1}} \bbh_l^{q}(\Delta)  \F_{l-1}^q\right\| \nonumber\\
    & \leq \sum_{q=1}^{F_{l-1}} \left\|    \bbh_l^{q}(\Delta_n) \bbx_{n,l-1}^q    - \samp_n^{\mathcal{X}}   \bbh_l^{q}(\Delta)  \F_{l-1}^q\right\|
 \end{align}
 The difference term in the last LHS of \eqref{proof_1} can be further decomposed for every $q=1,\dots,F_{l-1}$ as:
\begin{align}\label{proof_2}
   \nonumber   \|    \bbh_l^{q}(\Delta_n) & \bbx_{n,l-1}^q    - \samp_n^{\mathcal{X}}   \bbh_l^{q}(\Delta)  \F_{l-1}^q \| 
   \\ \nonumber&\leq \|
\bbh_l^{q}(\Delta_n) \bbx_{n,l-1}^q  - \bbh_l^{q}(\Delta_n) \samp_n^{\mathcal{X}} \F_{l-1}^q \\ &\qquad +\bbh_l^{q}(\Delta_n) \samp_n^{\mathcal{X}} \F_{l-1}^q  - \samp_n^{\mathcal{X}}   \bbh_l^{q}(\Delta)  \F_{l-1}^q
    \|\nonumber \\ \nonumber
   & \leq \left\|
    \bbh_l^{q}(\Delta_n) \bbx_{n,l-1}^q  - \bbh_l^{q}(\Delta_n) \samp_n^{\mathcal{X}} \F_{l-1}^q
    \right\|
  \\ &\qquad +
    \left\|
    \bbh_l^{q}(\Delta_n) \samp_n^{\mathcal{X}} \F_{l-1}^q  - \samp_n^{\mathcal{X}}   \bbh_l^{q}(\Delta)  \F_{l-1}^q
    \right\|
\end{align}
The first term of the last inequality in \eqref{proof_2} can be bounded as $\| \bbx_{n,l-1}^q - \samp_n^{\mathcal{X}}\F_{l-1}^q\|$ with the initial condition $\|\bbx_{n,0}^q - \samp_n^{\mathcal{X}} \F_0^q\|=0$ for $q = 1,\dots,F_0$. Denoting the second term with $D_{l-1}^n$,  and iterating the bounds derived above through layers and features, we obtain:
\begin{align}
 \nonumber \|\bPsi(\mathcal{H},\Delta_n,\samp_n^{\mathcal{X}} \F) - \samp_n^{\mathcal{X}} \bPsi(\mathcal{H},\Delta,\F)\|
 \leq
 \sum_{l=0}^L \prod\limits_{l'=l}^L F_{l'} D_l^n.
 \end{align}
 Therefore, we can focus on each difference term $D_l^n$ and omit the feature and layer indices to simplify notation. Considering that $\F$ and $\samp_n^{\mathcal{X}}\F$ are bandlimited, we can write the convolution operation as follows:
 \begin{align}
    & \nonumber\|\bbh(\Delta_n)\samp_n^{\mathcal{X}} \F - \samp_n^{\mathcal{X}}\bbh(\Delta) \F\| \nonumber \\
   & = \Bigg\| \sum_{i=1}^M \hat{h}(\lambda_i^n) \langle \samp_n^{\mathcal{X}}\F,\Phii^n \rangle_{\tm_n}\Phii^n \nonumber \\
   & \qquad \quad- \sum_{i=1}^M \hat{h}(\lambda_i)\langle \F,\Phii\rangle_{\tm} \samp_n^{\mathcal{X}} \Phii  \Bigg\| \nonumber
     \\ 
     &\nonumber \leq  \Bigg\| \sum_{i=1}^M \hat{h}(\lambda_i^n) \langle \samp_n^{\mathcal{X}}\F,\Phii^n \rangle_{\tm_n}\Phii^n \nonumber \\
     & \qquad \quad - \sum_{i=1}^M \hat{h}(\lambda_i) \langle \samp_n^{\mathcal{X}}\F,\Phii^n \rangle_{\tm_n}\Phii^n\Bigg\| \nonumber \\
     & \quad +\Bigg\| \sum_{i=1}^M \hat{h}(\lambda_i) \langle \samp_n^{\mathcal{X}} \F,\Phii^n \rangle_{\tm_n} \Phii^n \nonumber\\
     & \qquad \quad - \sum_{i=1}^M \hat{h}(\lambda_i) \langle \F,\Phii \rangle_{\tm} \samp_n^{\mathcal{X}} \Phii \Bigg\|,\label{eqn:conv-1}
 \end{align}
 where $M=\#\{\lambda_i\leq \lambda_M\}_i$ counts the number of eigenvalues within the bandwidth. 
The first term of the last bound in \eqref{eqn:conv-1} can be further bounded exploting the $C$-Lipschitz continuity of the frequency response function and the convergence in probability stated in \eqref{eqn:convergence_spectrum}: indeed, we can claim that for each eigenvalue $\lambda_i \leq \lambda_M$, for all $\epsilon_i>0$ and all $\delta_i>0$, there exists some $N_i$ such that for all $n>N_i$, we have:
\begin{gather}
 \label{eqn:eigenvalue}   \mathbb{P}(|\lambda_i^n-\lambda_i|\leq \epsilon_i)\geq 1-\delta_i,
 \end{gather}
Letting $\epsilon_i < \epsilon$ with $\epsilon > 0$, with probability at least $\prod_{i=1}^M(1-\delta_i) := 1-\delta$, we obtain:
\begin{align}
   &\left\| \sum_{i=1}^M (\hat{h}(\lambda_i^n) - \hat{h}(\lambda_i)) \langle \samp_n^{\mathcal{X}} \F,\Phii^n \rangle_{\tm_n} \Phii^n  \right\| \nonumber
   \\
   &\qquad \leq \sum_{i=1}^M |\hat{h}(\lambda_i^n)-\hat{h}(\lambda_i)| |\langle \samp_n^{\mathcal{X}} \F,\Phii^n \rangle_{\tm_n}| \|\Phii^n\| \nonumber\\
   &\qquad \leq \sum_{i=1}^M C |\lambda_i^n-\lambda_i| \|\samp_n^{\mathcal{X}} \F\| \|\Phii^n \|^2\leq MC\epsilon,
\end{align} 
for all $n>\max_i N_i := N$, where the first inequality is obtained applying the triangle inequality, the second inequality exploits the $C-$Lipschitz continuity of the frequency response, and the last inequality exploits \eqref{eqn:eigenvalue}.
The second term of the last bound in \eqref{eqn:conv-1} can be bounded eploiting the convergence of eigenvectors in \eqref{eqn:convergence_spectrum}. We start with
\begin{align}
   &\nonumber \left\| \sum_{i=1}^M \hat{h}(\lambda_i) \langle \samp_n^{\mathcal{X}} \F,\Phii^n \rangle_{\tm_n}\Phii^n - \sum_{i=1}^M \hat{h}(\lambda_i) \langle \F,\Phii \rangle_{\tm} \samp_n^{\mathcal{X}} \Phii \right\|\\
   &\leq \nonumber \left\|  \sum_{i=1}^M \hat{h}(\lambda_i) \left(\langle \samp_n^{\mathcal{X}} \F,\Phii^n\rangle_{\tm_n}\Phii^n  - \langle \samp_n^{\mathcal{X}} \F,\Phii^n \rangle_{\tm_n} \samp_n^{\mathcal{X}}\Phii\right)\right\|\\
   &\label{eqn:term1}+ \left\| \sum_{i=1}^M  \hat{h}(\lambda_i) \left(\langle \samp_n^{\mathcal{X}} \F,\Phii^n\rangle_{\tm_n} \samp_n^{\mathcal{X}}\Phii -\langle \F,\Phii\rangle_\tm \samp_n^{\mathcal{X}}\Phii \right) \right\|
\end{align}
From the convergence in probability stated in \eqref{eqn:convergence_spectrum}, we can claim that for some fixed eigenvector field $\Phii$,  for all $\epsilon_i>0$ and all $\delta_i>0$, there exists some $N_i$ such that for all $n>N_i$, we have
\begin{gather}
 \label{eqn:eigenfunction}    \mathbb{P}(\|\Phii^n - \samp_n^{\mathcal{X}}\Phii\|\leq \epsilon_i)\geq 1-\delta_i.
 \end{gather}
 Therefore, letting $\epsilon_i < \epsilon$ with $\epsilon > 0$, with probability at least $\prod_{i=1}^M(1-\delta_i) := 1-\delta$, for all $n> \max_i N_i := N$, the first term in \eqref{eqn:term1} can be bounded as
\begin{align}
&\nonumber \left\|  \sum_{i=1}^M \hat{h}(\lambda_i) \left(\langle \samp_n^{\mathcal{X}} \F,\Phii^n\rangle_{\tm_n}\Phii^n  - \langle \samp_n^{\mathcal{X}} \F,\Phii^n \rangle_{\tm_n} \samp_n^{\mathcal{X}}\Phii\right)\right\|\\
&\qquad \qquad \qquad \leq \sum_{i=1}^M \|\samp_n^{\mathcal{X}} \F\|\|\Phii^n - \samp_n^{\mathcal{X}}\Phii\|\leq M\epsilon,
\end{align}
considering the boundedness of frequency response function.
The second term in \eqref{eqn:term1} can be written as
\begin{align}
  \nonumber &\left\| \sum_{i=1}^M  \hat{h}(\lambda_i^n) \left(\langle \samp_n^{\mathcal{X}} \F,\Phii^n\rangle_{\tm_n} \samp_n^{\mathcal{X}}\Phii -\langle \F,\Phii\rangle_\ccalM \samp_n^{\mathcal{X}}\Phii \right) \right\| \\
   &\leq \sum_{i=1}^M|\hat{h}(\lambda_i^n)| \left|\langle \samp_n^{\mathcal{X}} \F,\Phii^n\rangle_{\tm_n}  -\langle \F,\Phii\rangle_\ccalM\right|\|\samp_n^{\mathcal{X}}\Phii\|.
\end{align}
Because $\{x_1, x_2,\cdots,x_n\}$ is a set of uniform sampled points from $\ccalM$, based on Proposition 11 in \cite{von2008consistency}, we can claim that 
\begin{equation}
    \lim_{n\to \infty} \mathbb{P}\left(\left|\langle \samp_n^{\mathcal{X}} \F,\Phii^n\rangle_{\tm_n}  -\langle \F,\Phii\rangle_\tm\right|\leq\epsilon \right)\geq 1-\delta,
\end{equation}
for all $\epsilon>0$ and $\delta>0$. Consider the boundedness of frequency response $|\hat{h}(\lambda)|\leq 1$ and the bounded energy of $\|\samp_n^{\mathcal{X}}\Phii\|$, we have for all $\epsilon>0$ and $\delta>0$:
\begin{align}
\lim_{n\to \infty}\mathbb{P}\Bigg(\Bigg\| \sum_{i=1}^M  \hat{h}(\lambda_i^n) &\bigg(\langle \samp_n^{\mathcal{X}} \F,\Phii^n\rangle_{\tm_n}  \nonumber \\
 &-\langle \F,\Phii\rangle_\tm\bigg)\samp_n^{\mathcal{X}}\Phii  \Bigg\|\leq M \epsilon\Bigg) \geq 1-\delta.
\end{align}

Combining all these results, we can claim that for all $\epsilon'>0$ and $\delta>0$, there exists some $N$, such that for all $n>N$ we have
\begin{equation}
    \mathbb{P}(\|\bbh(\Delta_n)\samp_n^{\mathcal{X}} f - \samp_n^{\mathcal{X}}\bbh(\Delta) \F\|\leq \epsilon')\geq 1-\delta.
\end{equation}

With $\lim\limits_{n\rightarrow \infty}D_l^n=0$ in high probability, this concludes the proof.

\subsection{Cellular Sheaves}
A cellular sheaf over an undirected graph consists of an assignment of a vector space to each node and edge in the graph and a map between these spaces for each incident node-edge pair. More formally, given an undirected graph $\mathcal{G} = (\mathcal{V}, \mathcal{E})$, with $|\mathcal{V}|=n$,  a cellular sheaf $\tm_n = (\mathcal{G}, \mathcal{F})$ over it consists of:
\begin{itemize}
    \item A vector space $\mathcal{F}(v)$ for each $v \in \mathcal{V}$. We refer to these vector spaces as nodes stalks.
    \item A vector space $\mathcal{F}(e)$ for each $e \in \mathcal{E}$. We refer to these vector spaces as edges stalks.
    \item A linear mapping $\mathbf{V}_{v,e}^T : \mathcal{F}(v) \rightarrow \mathcal{F}(e)$ for each incident $v \inc e$ node-edge pair. We refer to these mappings as restriction maps.
\end{itemize}
The space formed by all the spaces associated with the nodes of the graph is called
the space of 0-cochains (that we refer to as sheaf signals) $\mathcal{L}(\tm_n)$ (with a slight abuse of notation). The (non-normalized) Sheaf Laplacian of a sheaf $\tm_n$ is a linear mapping $\Delta_n:\mathcal{L}^2(\tm_n) \rightarrow \mathcal{L}^2(\tm_n)$ defined node-wise. In particular, given a sheaf signal $\mathbf{f}_n$, it holds: 
\begin{equation}
    (\Delta_n\mathbf{f}_n)(v) = \sum_{v,u\inc e}\mathbf{V}_{v,e}^T(\mathbf{V}_{v,e}\mathbf{f}_n(v) - \mathbf{V}_{u,e}\mathbf{f}_n(u)).
\end{equation}
The dimensions of the stalks and the restriction maps can be arbitrary defined: in this work, as the reader may have noticed at this point, our focus is on a specific class of cellular sheaves, called $\mathcal{O}(d)-$bundles (or orthogonal sheaves), i.e. sheaves with  orthogonal restriction maps and stalks with same dimension. In particular, it is now clear that, in our case, the sampled tangent bundle signals are sheaf signals, the dimension of the stalks is the estimated manifold dimension $\hd$, the restriction maps are given by the SVDs $\mathbf{M}_{i,j}$s and $\mathbf{V}^T_{i,j}$s  of the $\widetilde{\mathbf{O}}_{i,j}$s, and the (non-normalized) Sheaf Laplacian is given by the block matrix $\widetilde{\S} \in \mathbb{R}^{n\hd\times n\hd}$  with $\hd \times \hd$ blocks:
\begin{equation}
\widetilde{\S}_{i,j} = w_{i,j}\Oij
\end{equation}

An intuitive interpretation of cellular sheaves is given in \cite{hansen2020opinion} in terms of opinion dynamics. In this setting, the component $\mathbf{f}_n(v)$ of the sheaf signal $\mathbf{f}_n$ is the "private opinion" of node $v$, while $\mathbf{V}_{v,e}\mathbf{f}_n(v)$ describes how that private opinion publicly manifests  in the "discourse space"  $\mathcal{F}(e)$: in this sense, the Sheaf Laplacian applied to a  sheaf signal measures the aggregated "disagreement of opinions" at each node \cite{bodnar2022sheafdiff}.

\subsection{Consistency of Tangent Bundle Convolution} 
The tangent bundle convolution in Definition 1 is a generalization of the manifold convolution from \cite{wang2022convolution} and of the standard convolution on the real line. For the former case, the result is trivial, because manifold convolution is just the tangent bundle convolution in the case of scalar vector fields. For the latter case, consider the differential equation:
\begin{equation} \label{eqn:wave}
   \frac{\partial u(x,t)}{\partial t} = \frac{\partial}{\partial x} u(x,t) \text{,}
\end{equation}
which is a one-sided wave equation, thus it is not the exact
analogous of the diffusion equation in \eqref{diff_eq} for which we would require the second derivative to be used in the right side of \eqref{eqn:wave}. However, the important observation to make here is that the exponential of the derivative operator is a time shift operator so that we can write $u(x,t) = e^{-t\partial/\partial x}f(x) = f(x-t)$, where $f(x) = u(x,0)$; this is a known result and it holds because the operator $e^{-t\partial/\partial x}$ applied to $f$ evaluated in $x$ is equivalent to the Taylor Expansion of $f(x-t)$ around $x$. Another way of proving it is noticing that both $e^{t\partial/\partial x}f(x)$ and $f(x-t)$ are  solutions of \eqref{eqn:wave}. It then follows that Definition 1 particularized to \eqref{eqn:wave} yields the convolution definition:
\begin{equation}\label{eqn:conv-1d-1}
    g(x) = \int_{0}^\infty \tdh(t) e^{-t\partial/\partial x}f(x)\, \text{d}t.
         = \int_{0}^\infty \tdh(t) f(x-t) \,\text{d}t,
\end{equation}
that is the standard definition of time convolutions. 

The frequency representation result in Proposition 1 holds for \eqref{eqn:conv-1d-1} and it implies that standard convolutional filters in continuous time are completely characterized by the frequency response in Definition 3. The more standard definition of a filter's frequency response as the Fourier transform of the impulse response $\tdh(t)$  (as opposed to the  Laplace transform we use in Definition 3) suffices because complex exponentials $e^{jw}$ are an orthonormal basis of eigenfunctions of the derivative operator with associated eigenvalues $j\omega$. 
\end{document}